\documentstyle[aps,prl,multicol]{revtex}
\begin{document}
\draft
\title{Geometric phases for mixed states in
interferometry}
\author{Erik Sj\"{o}qvist$^{(1)}$, Arun K. Pati$^{(2,3)}$, 
Artur Ekert$^{(4)}$, Jeeva S. Anandan$^{(5)}$, Marie Ericsson$^{(1)}$, 
Daniel K.L. Oi$^{(4)}$, and Vlatko Vedral$^{(4)}$} 

\address{$^{(1)}$ Department of Quantum Chemistry, Uppsala University,
Box 518, Se-751 20 Sweden}

\address{$^{(2)}$ School of Informatics, University of Wales,
Bangor LL 57 1UT, UK}
\address{$^{(3)}$ Theoretical Physics Division, 5th Floor, C. C.,
BARC, Mumbai-400085, India}
\address{$^{(4)}$ Centre for Quantum Computation, University 
of Oxford, Clarendon Laboratory, Parks Road, Oxford OX1 3PU, 
UK}
\address{$^{(5)}$ Department of Physics and Astronomy,
University of South Carolina, Columbia, SC 29208, USA}
\maketitle
\begin{abstract}
We provide a physical prescription based on interferometry 
for introducing the total phase of a mixed state undergoing 
unitary evolution, which has been an elusive concept in the 
past. We define the parallel transport condition that 
provides a connection-form for obtaining the geometric phase 
for mixed states. The expression for the geometric phase for 
mixed state reduces to well known formulas in the pure state 
case when a system undergoes noncyclic and unitary quantum 
evolution.
\end{abstract}
\begin{multicols}{2}

When a pure quantal state undergoes cyclic evolution 
the system returns to its original state but may acquire 
a phase factor of purely geometric origin. Though this 
was realized in the adiabatic context \cite{mvb}, the 
nonadiabatic generalization was found in \cite{aa}. 
Based on Pancharatnam's \cite{pancharatnam56} earlier 
work, this concept was generalized to noncyclic 
evolutions of quantum systems \cite{sb}. Subsequently, 
the kinematic approach \cite{ms} and gauge potential 
description \cite{akp,akp1} of geometric phases for 
noncyclic and non-Schr\"{o}dinger evolutions were 
provided. The adiabatic Berry phase and Hannay angle 
for open paths were introduced \cite{akp2} and discussed 
\cite{es}. The noncyclic geometric phase has been 
generalized to non-Abelian cases \cite{ali}. Applications 
of geometric phase have been found in molecular dynamics 
\cite{sh}, response function of many-body system \cite{jp,ap}, 
and geometric quantum computation \cite{jvec,ekert00}.
Noncyclic geometric phase for entangled states has also been
studied \cite{erik}. In all these developments the geometric
phase has been discussed only for {\em pure} states. However, in some 
applications, in particular geometric fault tolerant
quantum computation \cite{jvec,ekert00}, we are primarily 
interested in mixed state cases. Uhlmann was probably the 
first to address the issue of mixed state holonomy, but as 
a purely mathematical problem \cite{uhlmann86,uhlmann91}. 
In contrast, here we provide a new formalism of geometric 
phase for mixed states in the experimental context of 
quantum interferometry. 

The purpose of this Letter is to provide an {\em operationally
well defined} notion of phase for unitarily evolving mixed 
quantal states in interferometry, which has been an elusive 
concept in the past. This phase fulfills two
central properties that makes it a natural generalization 
of the pure case: (i) it gives rise to a linear shift of 
the interference oscillations produced by a variable $U(1)$ 
phase, and (ii) it reduces to the Pancharatnam connection 
\cite{pancharatnam56} for pure states. We introduce the notion 
of parallel transport based on our defintion of total phase. 
We moreover introduce a concept of geometric phase for unitarily 
evolving mixed quantal states. This geometric phase reduces to 
the standard geometric phase \cite{ms,akp,akp1} for pure states 
undergoing noncyclic unitary evolution. 

{\it Mixed states, phases and interference:}
Consider a conventional Mach-Zehnder interferometer 
in which the beam-pair spans a two dimensional  
Hilbert space $\tilde{{\cal H}} = \{ |\tilde{0}\rangle , 
|\tilde{1}\rangle \}$. The state vectors $|\tilde{0}\rangle$ 
and $|\tilde{1}\rangle$ can be taken as wave packets that 
move in two given directions defined by the geometry of 
the interferometer. In this basis, we may represent  
mirrors, beam-splitters and relative $U(1)$ phase shifts 
by the unitary operators 
\begin{eqnarray}
\tilde{U}_{M} & = & \left( 
\begin{array}{rr} 0 & 1 \\ 
 1 & 0 \end{array} \right) , \, \, 
\tilde{U}_{B} = \frac{1}{\sqrt{2}} \left( 
\begin{array}{rr} 1 & 1 \\ 
 1 & -1 \end{array} \right) , 
\nonumber \\ 
\tilde{U}(1) & = & \left( 
\begin{array}{cr} e^{i\chi} & 0 \\ 
 0 & 1 \end{array} \right) ,  
\label{eq:spatialunitary} 
\end{eqnarray}
respectively. An input pure state $\tilde{\rho}_{{\text{in}}} 
= | \tilde{0} \rangle \langle \tilde{0} |$ of the interferometer 
transforms into the output state 
\begin{eqnarray}
\tilde{\rho}_{{\text{out}}} & = & 
\tilde{U}_{B} \tilde{U}_{M} \tilde{U}(1)
\tilde{U}_{B} \tilde{\rho}_{{\text{in}}} \tilde{U}_{B}^{\dagger} 
\tilde{U}^{\dagger}(1) \tilde{U}_{M}^{\dagger} 
\tilde{U}_{B}^{\dagger} \nonumber \\ 
 & = & \frac{1}{2} \left( \begin{array}{cc} 
1 + \cos \chi & i \sin \chi  \\ 
-i \sin \chi & 1 - \cos \chi  \end{array} 
\right) 
\end{eqnarray}
that yields the intensity along $|\tilde{0}\rangle$ as 
$I \propto 1+\cos \chi$. Thus the relative $U(1)$ phase $\chi$
could be observed in the output signal of the interferometer. 

Now assume that the particles carry additional internal 
degrees of freedom, e.g., spin. This internal spin space 
${\cal H}_{{\text{i}}} \cong {\cal C}^N$ is spanned by the vectors  
$|k\rangle$, $k=1,2, \ldots N$, chosen so that the 
associated density operator is initially diagonal 
\begin{equation}
\rho_{0} = \sum_{k} w_{k} 
|k\rangle \langle k| 
\label{eq:generalinternal}
\end{equation}
with $w_{k}$ the classical probability to find a 
member of the ensemble in the pure state $|k\rangle$. The density 
operator could be made to change inside the interferometer   
\begin{equation} 
\rho_{0} \longrightarrow  U_{{\text{i}}} 
\rho_{0} U_{{\text{i}}}^{\dagger}  
\end{equation}
with $U_{{\text{i}}}$ a unitary transformation acting only on 
the internal degrees of freedom. Mirrors and beam-splitters 
are assumed to leave the internal state unchanged so that 
we may replace  $\tilde{U}_{M}$ and $\tilde{U}_{B}$ by 
${\bf U}_{M} = \tilde{U}_{M}\otimes 1_{{\text{i}}}$ and 
${\bf U}_{B} = \tilde{U}_{B}\otimes 1_{{\text{i}}}$, 
respectively, $1_{{\text{i}}}$ being the internal unit 
operator. Furthermore, we introduce the unitary transformation 
\begin{equation}
{\bf U} = \left( \begin{array}{rr} 
 0 & 0 \\ 0 & 1 \end{array} \right) \otimes U_{{\text{i}}} + 
\left( \begin{array}{cr} 
 e^{i\chi} & 0 \\ 0 & 0 \end{array} \right) \otimes 
1_{{\text{i}}} . 
\label{eq:unitary}
\end{equation} 
The operators ${\bf U}_{M}$, ${\bf U}_{B}$, and 
${\bf U}$ act on the full Hilbert space 
$\tilde{{\cal H}} \otimes {\cal H}_{{\text{i}}}$. 
${\bf U}$ correponds to the application of $U_{{\text{i}}}$ 
along the $|\tilde{1}\rangle$ path and the $U(1)$ phase 
$\chi$ similarly along $|\tilde{0}\rangle$. We shall use 
${\bf U}$ to generalize the notion of phase to unitarily 
evolving mixed states.  

Let an incoming state given by the density matrix 
$\varrho_{{\text{in}}} = 
\tilde{\rho}_{{\text{in}}} \otimes \rho_{0} = 
|\tilde{0} \rangle \langle \tilde{0} | \otimes \rho_{0}$
be split coherently by a beam-splitter and recombine 
at a second beam-splitter after being reflected by two 
mirrors. Suppose that ${\bf U}$ is applied between the 
first beam-splitter and the mirror pair. The incoming 
state transforms into the output state 
\begin{equation}
\varrho_{{\text{out}}} = 
{\bf U}_{B} {\bf U}_{M} {\bf U} {\bf U}_{B} 
\varrho_{{\text{in}}} {\bf U}_{B}^{\dagger} 
{\bf U}^{\dagger} {\bf U}_{M}^{\dagger} 
{\bf U}_{B}^{\dagger} . 
\label{eq:outputmatrix}
\end{equation}
Inserting Eqs.~(\ref{eq:spatialunitary}) and (\ref{eq:unitary}) 
into Eq.~(\ref{eq:outputmatrix}) yields 
\begin{eqnarray}
\varrho_{{\text{out}}} & = & \frac{1}{4} \left[ 
\left( \begin{array}{rr} \phantom{,}1 & \phantom{-}1 \\ 
1 & 1 \end{array} \right) 
\otimes U_{{\text{i}}} \rho_{0} 
U_{{\text{i}}}^{\dagger} + 
\left( \begin{array}{rr} 1 & -1 \\ -1 & 1 
\end{array} \right) \otimes \rho_{0} \right. 
\nonumber \\ & & + e^{i\chi} 
\left( \begin{array}{rr} 1 & 1 \\ -1 & -1 \end{array} 
\right) \otimes \rho_{0} U_{{\text{i}}}^{\dagger} 
\nonumber \\ & & \left. + 
e^{-i\chi} \left( \begin{array}{rr} \phantom{,}1 & -1 \\ 
1 & -1 \end{array} 
\right) \otimes U_{{\text{i}}} \rho_{0} 
\right] . 
\end{eqnarray} 
The output intensity along $|\tilde{0}\rangle$ is  
\begin{eqnarray} 
I & \propto & {\text{Tr}} \left( U_{{\text{i}}} 
\rho_{0} U_{{\text{i}}}^{\dagger} + 
\rho_{0} + e^{-i\chi} U_{{\text{i}}} 
\rho_{0} + e^{i\chi} \rho_{0} 
U_{{\text{i}}}^{\dagger} \right) 
\nonumber \\ 
 & \propto & 1 + |{\text{Tr}} \left( U_{{\text{i}}} 
\rho_{0} \right)| \cos \left[ \chi - 
\arg {\text{Tr}} \left( U_{{\text{i}}} 
\rho_{0} \right) \right] , 
\label{eq:outputintensity}
\end{eqnarray}
where we have used ${\text{Tr}} 
( \rho_{0} U_{{\text{i}}}^{\dagger} ) = 
\left[ {\text{Tr}} \left( U_{{\text{i}}} \rho_{0} 
\right) \right]^{\ast}$. 

The important observation from Eq.~(\ref{eq:outputintensity})
is that the interference oscillations produced by 
the variable $U(1)$ phase $\chi$ {\em is shifted by 
$\phi = \arg {\text{Tr}} \left( U_{{\text{i}}} 
\rho_{0} \right)$ for any internal input state 
$\rho_{0}$,} be it mixed or pure. This phase difference 
reduces for pure states 
$\rho_{0} = |\psi_{0} \rangle \langle \psi_{0}|$ 
to the Pancharatnam phase difference between $ U_{{\text{i}}} 
|\psi_{0} \rangle$ and   
$|\psi_{0} \rangle$. These two latter facts are 
the central properties for $\phi$ being a natural 
generalization of the pure state phase. Moreover 
the visibility of the interference pattern is 
$\nu = |{\text{Tr}} \left( U_{{\text{i}}} \rho_{0} 
\right)| \geq 0$, which reduces to the expected $\nu = 
|\langle \psi_{0} |U_{{\text{i}}}|\psi_{0} \rangle|$ 
for pure states. 

The output intensity in Eq.~(\ref{eq:outputintensity}) 
may be understood as an incoherent weighted average of 
pure state interference profiles as follows. The state 
$k$ gives rise to the interference profile 
\begin{equation}
I_{k} \propto 1+\nu_{k}\cos \left[ \chi -\phi_{k} \right] , 
\end{equation}
where $\nu_{k} = | \langle k|U_{{\text{i}}}|k \rangle |$ 
and $\phi_{k} = \arg \langle k|U_{{\text{i}}}|k \rangle$. 
This yields the total output intensity 
\begin{equation}
I=\sum_{k} w_{k} I_{k} \propto 
1+\sum_{k} w_{k} \nu_{k}\cos \left[ \chi -\phi_{k} \right] , 
\label{eq:average}
\end{equation}
which is the incoherent classical average of the above 
single-state interference profiles weighted by the 
corresponding probabilities $w_{k}$. 
Eq.~(\ref{eq:average}) may be written in the desired 
form $1+\tilde{\nu} \cos (\chi -\tilde{\phi})$ by making 
the identifications 
\begin{eqnarray}
\tilde{\phi} & = & \arg \left( \sum_{k} w_{k} 
\nu_{k} e^{i\phi_{k}} \right) = \arg {\text{Tr}} \left( 
U_{{\text{i}}} \rho_{0} \right) = \phi , \nonumber \\ 
\tilde{\nu} & = & \left| \sum_{k} w_{k} 
\nu_{k} e^{i\phi_{k}} \right| = 
|{\text{Tr}} \left( U_{{\text{i}}} \rho_{0} \right) | = 
\nu .  
\label{eq:totalaverage}
\end{eqnarray}

{\it Parallel transport condition and geometric phase:}
Consider a continuous unitary transformation of the mixed 
state given by $\rho(t) =  U(t) \rho_{0} U^{\dagger}(t)$. 
(From now on, we omit the subscript ``i'' of $U$.) We say 
that the state of the system $\rho(t)$ acquires a phase 
with respect to $\rho_{0}$ if $\arg {\text {Tr}}[ U(t) \rho_{0}]$ 
is nonvanishing. Now if we want to parallel transport a 
mixed state $\rho(t)$ along an arbitrary path, then at each 
instant of time the state must be {\em in-phase} with the 
state at an infinitesimal time. The state at time $t+dt$ 
is related to the state at time $t$ as $\rho(t+dt) = 
U(t+dt)U^{\dagger}(t) \rho(t) U(t)U^{\dagger}(t+dt)$. 
Therefore, the phase difference between $\rho(t)$ and 
$\rho(t+dt)$ is $\arg {\text{Tr}}[\rho(t) U(t+dt)U^{\dagger}(t) ]$. 
We can say $\rho(t)$ and $\rho(t+dt)$ are in phase if 
${\text{Tr}}[\rho(t) U(t+dt)U^{\dagger}(t) ]$ is {\em real 
and positive}. This condition can be regarded as a generalization 
of Pancharatnam's connection from pure to mixed states. However, 
from normalization and Hermiticity of $\rho(t)$ it follows 
that ${\text{Tr}}[\rho(t) {\dot U(t)}U^{\dagger}(t) ]$ is 
purely imaginary. Hence the above mixed state generalization 
of Pancharatnam's connection can be met only when  
\begin{equation}
{\text{Tr}}[\rho(t) {\dot U(t)} U^{\dagger}(t) ] = 0.
\label{eq:parallelcondition}
\end{equation}
This is the parallel transport condition for mixed 
states undergoing unitary evolution. On the space of 
density matrices the above condition can be translated 
to ${\text{Tr}}[\rho~dU~U^{\dagger}] = 0$, where $d$ 
is the exterior derivative on the space of density 
operators. However, $\rho(t)$ determines the $N \times N$ 
matrix $U(t)$ ($N$ being the dimension of the Hilbert 
space) up to $N$ phase factors, and the single condition 
Eq. (\ref{eq:parallelcondition}) while necessary is not 
sufficient to determine $U(t)$. These $N$ phase factors 
are fixed by the $N$ parallel transport conditions  
\begin{equation}
\langle k(t) | \dot{U} (t) U^{\dagger} (t) | k(t) \rangle = 0 , 
\, \, \, k=1,2,\ldots N , 
\label{eq:pureparallelcondition}
\end{equation}
where the $| k(t) \rangle$'s are orthonormal eigenstates of
$\rho(t)$. These are sufficient to determine the parallel 
transport operator $U(t)$ if we are given a non-degenerate 
density matrix $\rho(t)$.

The parallel transport condition for a mixed state provides 
us a {\em connection} in the space of density operators which 
can be used to define the geometric phase. Thus a
mixed state can acquire pure geometric phase if it 
undergoes parallel transport along an arbitrary curve. One 
can check that if we have a pure state density operator 
$\rho(t) = |\psi (t) \rangle \langle \psi (t) \arrowvert$ 
then the parallel transport condition 
Eq.~(\ref{eq:parallelcondition}) reduces to 
$\langle \psi (t) \arrowvert {\dot \psi} (t) \rangle = 0$
as has been discussed in \cite{aa,sb,ms,akp,akp1,jsa1,jsa} which is
both necessary and sufficient.

Now we can define a geometric phase for mixed state evolution. 
Let the state trace out an open unitary curve $\Gamma : t \in 
[0,\tau] \longrightarrow \rho (t) = U(t) \rho_{0} U^{\dagger} (t)$ 
in the space of density operators with ``end-points'' $\rho (0) = 
\rho_{0}$ and $\rho (\tau)$, where $U(t)$ satisfies Eq. 
(\ref{eq:parallelcondition}). The evolution need not be cyclic, 
i.e. $\rho (\tau) \neq \rho_{0}$. We can naturally assign a 
geometric phase $\gamma_{g} [\Gamma]$ to this curve once we notice that
the dynamical phase vanishes identically. The
dynamical phase is the time integral of the average of Hamiltonian
and can be defined as
\begin{eqnarray}
\gamma_{d} =
- \frac{1}{\hbar} \int_{0}^{\tau}~dt~{\text{Tr}} [\rho(t) H(t)] 
\nonumber \\
= -i \int_{0}^{\tau}~dt~{\text{Tr}} [
\rho_{0} U^{\dagger} (t) \dot{U} (t) ].
\end{eqnarray}
Since the density matrix undergoes parallel transport evolution 
the dynamical phase vanishes identically. Moreover, the parallel 
transport operator $U(t)$ should fulfill the stronger condition 
Eq. (\ref{eq:pureparallelcondition}). Thus the geometric phase 
for a mixed state is defined as 
\begin{eqnarray}
\gamma_{g} [\Gamma] & = & \phi = 
\arg {\text{Tr}} [ \rho_{0} U(t) ] = 
\arg \left( \sum_k w_k \nu_{k} e^{i\beta_k} \right) , 
\label{eq:gpmixed}
\end{eqnarray}
where $\exp (i \beta_k)$ are geometric phase factors associated 
with the individual pure state paths in the given ensemble. The 
above geometric phase can be given a gauge potential description 
such that the line integral will give the open path geometric 
phase for mixed state evolution. Indeed the mixed state holonomy 
can be expressed as
\begin{eqnarray} 
\gamma_{g} [\Gamma] & = & \int~dt~ i {\text{Tr}} [ \rho_{0}  
W^{\dagger}(t) {\dot W}(t) ] 
\nonumber \\ 
 & = & \int_{\Gamma} i {\text{Tr}} 
\left[ \rho_{0} W^{\dagger} dW \right] = 
\int_{\Gamma} d \Omega,
\label{eq:gppotential}
\end{eqnarray} 
where 
\begin{equation}
W(t) = \frac{{\text{Tr}} [ \rho_{0} U^{\dagger}(t) ]}
{|{\text{Tr}} [ \rho_{0} U^{\dagger}(t) ] |} U(t) . 
\end{equation}
and $U(t)$ satisfies (\ref{eq:pureparallelcondition}).
The quantity $\Omega = i {\text{Tr}} \left[ \rho_{0}
W^{\dagger} dW \right]$ can be regarded as a gauge 
potential on the space of density operators pertaining 
to the system. 

The geometric phase defined above is manifestly gauge invariant, 
does not depend explicitly on the dynamics but it depends only 
on the geometry of the open unitary path $\Gamma$ in the space 
of density operators pertaining to the system. It is also 
independent of the rate at which the system is transported 
in the quantum state space. The geometric phase Eq. 
(\ref{eq:gppotential}) can also be expressed in terms 
of an average connection form
\begin{eqnarray} 
\gamma_{g} [\Gamma] & = & 
\int_{\Gamma} \sum_k w_k i \langle \chi_k | d \chi_k \rangle 
= \int_{\Gamma} \sum_k w_k \Omega_k,  
\end{eqnarray} 
where $\Omega_k$ is connection-form and $|\chi_k(t) \rangle
=W(t)| k \rangle$ is the
``reference-section'' for $k$th component in the ensemble.
To be sure, what we have defined is consistent with known 
results, we can check that this expression reduces to the 
standard geometric phase \cite{ms,akp,akp1} 
\begin{eqnarray}
\gamma_{g} [\Gamma] & = & 
\arg \langle \psi (0) |\psi (\tau) \rangle =
\int_{0}^{\tau}~dt~i \langle \chi(t) | \dot{\chi} (t) \rangle
\label{eq:gppure}
\end{eqnarray}
for a pure state $\rho (t) = |\psi (t) \rangle \langle 
\psi (t)|$ when it satisfies parallel transport condition.
Here, $|\chi(t) \rangle$ is a reference state,
which gives the generalised connection one-form \cite{akp,akp1}.

{\it Purification:} 
An alternative approach to the above results is given 
by lifting the mixed state into a purified state 
$|\Psi \rangle$ by attaching an ancilla. We can imagine 
that any mixed state can be obtained by tracing out some 
degrees of freedom of a larger system which was in a pure 
state
\begin{equation}
|\Psi \rangle=
\sum_k \sqrt{w_{k}} |k\rangle_s |k \rangle_a ,
\end{equation} 
where $|k \rangle_a$ is a basis in an auxiliary Hilbert space, 
describing everything else apart from the spatial and the spin
degrees of freedom. The existence of the above purification 
requires that the dimensionality of the auxiliary Hilbert space 
is at least as large as that of the internal Hilbert space. 
If $|\Psi \rangle$ is the state at time $t=0$ and it is 
transformed to $|\Psi(t) \rangle$ by a local unitary operator 
$U(t)= U_s(t) \otimes I_a$ then
\begin{equation}
|\Psi(t) \rangle=\sum_k \sqrt{w_{k}} U_s(t)|k\rangle_s |k \rangle_a .
\end{equation}
The inner-product of initial and final state 
\begin{equation}
\langle \Psi(0)|\Psi(t) \rangle = 
\sum_k w_{k} \langle k|U(t)|k \rangle = {\text{Tr}} (U(t)\rho_{0}) 
\label{eq:purification}
\end{equation}
gives the full description of the modified interference. 
Indeed by comparing Eqs.~(\ref{eq:outputintensity}) and 
(\ref{eq:purification}), we see that $\arg \langle \Psi(0)|
\Psi(t) \rangle$ is the phase shift and $|\langle \Psi(0)|
\Psi(t) \rangle |$ is the visibility of the output intensity 
obtained in an interferometer. 

The parallel transport condition, given by 
Eq.~(\ref{eq:parallelcondition}), follows immediately from 
the pure state case when applied to any purification of 
$\rho_{0}$, i.e.
\begin{eqnarray} 
0 & = & \langle \Psi(t) |\dot{\Psi}(t) \rangle = 
\sum_k w_k \langle k| U^{\dagger}(t){\dot U(t)} |k\rangle 
\nonumber \\ 
 & = & {\text{Tr}} [ \rho_{0} U^{\dagger}(t){\dot U(t)} ] = 
{\text{Tr}} [ \rho (t){\dot U(t)}U^{\dagger}(t) ] .
\end{eqnarray} 
Thus a parallel transport of a density operator $\rho(t)$ 
amounts to a parallel transport of any of its purifications. 

{\it Example:} 
Consider a qubit (a spin-$\case{1}{2}$ particle) whose 
density matrix can be written as
\begin{equation}
\rho = \frac{1}{2} (1+r\hat{{\bf r}} \cdot 
{\mbox{\boldmath $\sigma$}}) , 
\label{eq:2levelstate}
\end{equation}
where $\hat{{\bf r}}$ is a unit vector and $r$ is constant 
for unitary evolution. The pure states $r = 1$ define the 
unit Bloch sphere containing the mixed states $r < 1$. 
Suppose that during the time evolution $\hat{{\bf r}}$ traces 
out a curve on the Bloch sphere that subtends a geodesically 
closed solid angle $\Omega$ \cite{jsa1}. The two pure states 
$|\pm ;\hat{{\bf r}} \cdot {\mbox{\boldmath $\sigma$}} \rangle$ 
acquire noncyclic geometric phase $\mp \Omega /2$ and identical 
visibility $\nu_{+} = \nu_{-} \equiv \eta$. Using 
Eq.~(\ref{eq:gpmixed}) we obtain the geometric phase 
for $\Gamma$ 
\begin{equation}
\phi = \gamma_{g} [\Gamma] = 
-\arctan \left( r \tan \frac{\Omega}{2} \right) . 
\label{eq:su2gp}
\end{equation}
The visibility $\nu = |{\text{Tr}} 
\left( U \rho_{0}\right) |$ is given by
\begin{eqnarray}
\nu & = & \eta \sqrt{\cos^{2} \frac{\Omega}{2} + r^{2}
\sin^{2} \frac{\Omega}{2}} .  
\label{eq:su2visibility}
\end{eqnarray}
For cyclic evolution we have $\eta = 1$ but the mixed state 
$\nu \neq 1$ due to the square root factor on the 
right-hand side of Eq.~(\ref{eq:su2visibility}). Moreover 
Eqs.~(\ref{eq:su2gp}) and (\ref{eq:su2visibility}) 
reduce to the usual expressions for pure states 
$\phi = -\Omega /2$ and $\nu = \eta$ by letting $r=1$. 

In the case of maximally mixed states $r=0$ we obtain 
$\phi = \arg \cos (\Omega /2)$ and $\nu = |\cos (\Omega /2)|$.
Thus the output intensity for such states is 
\begin{eqnarray}
I & \propto & 1 + |\cos \frac{\Omega}{2}| \cos \left[\chi - 
\arg \cos \frac{\Omega}{2} \right] 
\nonumber \\ 
 & = & 1 + \cos \frac{\Omega}{2} 
\cos \chi . 
\label{eq:2levelunpol}
\end{eqnarray}
Early experiments \cite{rauch75,werner75,klein76} to test 
the $4\pi$ symmetry of spinors utilized unpolarized neutrons. 
Eq.~(\ref{eq:2levelunpol}) show that in these experiments 
the sign change for $\Omega = 2\pi$ is a consequence of the 
phase shift $\phi = \arg \cos \pi = \pi$. 

Note that $\gamma_{g} [\Gamma]$ in Eq.~(\ref{eq:su2gp}) 
equals the geodesically closed solid angle on the Poincar\'{e} 
sphere iff $r=1$. In the mixed state case the geometric phase 
factor is weighted average of the solid angles subtended by 
the two pure state paths on the Bloch sphere.

In conclusion, we have provided a physical prescription 
based on interferometry for introducing a concept of 
total phase for mixed states undergoing unitary evolution. 
We have provided the necessary and sufficient condition for
parallel transport of a mixed state and introduced a concept of
geometric phase for mixed states when it undergoes parallel transport.
This reduces to known formulas for pure
state case when the system follows a noncyclic and unitary
quantum evolutions. We have also provided a gauge potential for
noncyclic evolutions of mixed states whose line integral
gives the geometric phase. We hope this will lead to 
experimental test of geometric phases for mixed states 
and further generalization of it to nonunitary and 
nonlinear evolutions.

The work by E.S. was financed by the Swedish Natural 
Science Research Council (NFR). A.K.P. acknowledges 
EPSRC for financial support and UK Quantum Computing 
Network for supporting his visit to Centre for Quantum 
Computation, Oxford. J.S.A. thanks Y. Aharonov and 
A. Pines for useful discussions and NSF and ONR grants 
for financial support.M.E. acknowledges financial support 
from the European Science Foundation. D.K.L.O. acknowledges 
financial support from CESG.

\end{multicols}
\end{document}